\documentclass[usenatbib]{mn2e}
\usepackage{amssymb}
\usepackage{times}
\usepackage{epsfig}
\usepackage{colordvi}
\usepackage{color}

\newcommand{\msun}{\ensuremath{{\rm M}_{\odot}}}
\newcommand{\xte}{{\it RXTE}}
\newcommand{\igr}{IGR17511}
\newcommand{\xmm}{{\it XMM}}
\newcommand{\swift}{{\it Swift}}
\newcommand{\kte}{\ensuremath{T_\mathrm{e}}}
\newcommand{\ktseed}{\ensuremath{T_\mathrm{seed}}}
\newcommand{\tin}{\ensuremath{T_\mathrm{in}}}
\newcommand{\ktbb}{\ensuremath{T_\mathrm{bb}}}
\newcommand{\ktdisk}{\ensuremath{T_\mathrm{in}}}

\newcommand{\rstar}{\ensuremath{R_\ast}}
\newcommand{\nh}{\ensuremath{N_\mathrm{H}}}
\newcommand{\mstar}{\ensuremath{M_\ast}}

\newcommand{\rin}{\ensuremath{R_\mathrm{in}}}

\newcommand{\be}{\begin{eqnarray}}
\newcommand{\ee}{\end{eqnarray}}
\newcommand{\refl}{{\Re}}

\def\aapr{A\&A~Rev.}          
\def\apj{ApJ}
\def\apjl{ApJ}
\def\apjs{ApJS}
\def\aap{A\&A}
\def\mnras{MNRAS}
\def\nat{Nat}              
\def\pasj{PASJ}

\begin{document}
\title[IGR J17511--3057 in 2009]{The 2009 outburst of  accreting millisecond pulsar IGR~J17511--3057
as observed by \textit{SWIFT} and \textit{RXTE}}

\author[A. Ibragimov, J. J. E. Kajava and J. Poutanen]
{Askar Ibragimov,$^{1,2}$\thanks{E-mail: askar.ibragimov@oulu.fi, jari.kajava@oulu.fi, juri.poutanen@oulu.fi}
Jari J. E. Kajava$^2$ and Juri Poutanen$^2$\\
$^1$Sabanc{\i} University,  Orhanl{\i}-Tuzla, Istanbul, 34956, Turkey\\
$^2$Astronomy Division, Department of Physics, PO Box 3000, FIN-90014 University of Oulu, Finland}

\pagerange{\pageref{firstpage}--\pageref{lastpage}}
\pubyear{2011}
\date{Accepted 2011 April 01. Received 2011 March 18; in original form 2011 February 09}

\maketitle
\label{firstpage}
\begin{abstract}
The twelfth accretion-powered millisecond pulsar, IGR J17511--3057, was discovered in September 2009.  In this work we study its spectral and timing properties during the 2009 outburst based on \textit{Swift} and \textit{RXTE} data.
Our  spectral analysis of the source indicates only slight spectral shape evolution during the entire outburst.
The equivalent width of the iron line and the apparent area of the blackbody emission associated with the hotspot at the stellar surface
both decrease significantly during the outburst. This is consistent with a gradual receding of the accretion disc as the accretion rate drops.
The pulse profile analysis shows absence of dramatic shape evolution with a moderate decrease in pulse amplitude.
This behaviour might result from a movement  of the accretion column footprint towards the magnetic pole as the disc retreats.
The time lag between the soft and the hard energy pulses  increase by a factor of two during the outburst.
A physical displacement of the centroid of the accretion shock relative  to the blackbody spot
or changes in the emissivity pattern of the Comptonization component related to the variations of the accretion column structure
could cause this evolution.
We have found that IGR J17511--3057 demonstrates outburst stages similar to those seen in SAX J1808.4--3658.
A transition from the ``slow decay'' into  the ``rapid drop'' stage, associated with the dramatic flux decrease, is also accompanied by a
pulse phase shift which could result from an appearance of the secondary spot due to the increasing inner disc radius.

\end{abstract}
\begin{keywords}
accretion, accretion discs -- methods: data analysis -- pulsars: individual: IGR J17511--3057 -- X-rays:binaries
\end{keywords}

\section{Introduction}
Accretion-powered millisecond pulsars (AMPs) are neutron stars (NS) that experience transient accretion episodes and show millisecond pulsations corresponding  to the stellar rotational period.
Presently, 13 objects of this class are known \citep[see reviews by][]{P06,W06,2010arXiv1007.1108P}.
The pulsations arise because the NS magnetic field channels the accretion flow on to the NS magnetic poles.
Such an accretion flow produces close to sinusoidal pulse profiles in most AMPs, but for several of them a strong evolution and peculiar double-peaked pulse shapes are observed (e.g. \citealt{2008ApJ...675.1468H, 2009MNRAS.400..492I}).

The energy spectra of AMPs contain 0.5--1 keV blackbody emission from the neutron star surface and a Comptonization component
dominating at higher energies and associated probably  with the accretion shock
(e.g., \citealt*{2002MNRAS.331..141G}; \citealt{2003MNRAS.343.1301P, 2005A&A...444...15F, 2007A&A...464.1069F}).
In addition, emission from the accretion disc at soft energies ($\lesssim 2$ keV) has also been detected in {\it XMM-Newton} observations of XTE J1751--305 \citep{2005MNRAS.359.1261G}, XTE J1807--294 \citep{2005A&A...436..647F}, SAX J1808.4--3658 \citep{2009MNRAS.396L..51P} and recently in IGR J17511--3057 \citep[hereafter \igr, ][]{2010MNRAS.407.2575P}. The reflection component of moderate amplitude is also detected in AMPs \citep{2005MNRAS.359.1261G, 2009MNRAS.400..492I}.

The pulse profiles are often rather sinusoidal with a slight skewness \citep{P06} and show clear energy dependence.
Pulses  at soft energies peak at a later phase resulting in ``soft lags'', as first was seen in SAX J1808.4--3658 \citep*{1998ApJ...504L..27C}.
The origin of the soft lag is most likely related to the different angular emission patterns of the blackbody and Comptonized components, which naturally explains why  the phase lags seem to saturate at $\sim 7$--$10$ keV, where the emission of the blackbody component becomes negligible \citep{2002MNRAS.331..141G, 2003MNRAS.343.1301P, 2005MNRAS.359.1261G}.
Some AMPs however show also a (not fully understood) decrease in the lag above $\sim 10$ keV (seen in IGR J00291+5934 as indicated by \citealt{2005ApJ...622L..45G} and \citealt{2005A&A...444...15F}; \igr\ might also have this decrease above $\sim 20$ keV, see \citealt{2010arXiv1012.0229F}). Information about this lag can be used to study the properties of the  accretion shock and the structure of the hotspot.

\subsection{IGR J17511--3057}

\igr\ was discovered on 2009 September 12 (MJD 55087) by \textit{INTEGRAL}  observatory \citep{2009ATel.2196....1B} during the Galactic bulge monitoring program \citep{2007A&A...466..595K}.
The 245 Hz pulsations were detected by \xte\ \citep{2009ATel.2197....1M} confirming the AMP nature of \igr.
A {\it Chandra}/HETG observation provided the source position of (J2000) RA=$17^{\rm h}51^{\rm m}08\fs66$, Dec=$-30\degr 57\arcmin 41\farcs0$ ($1\sigma$ error of $0\farcs6$, \citealt{2009ATel.2215....1N}).
A near infrared counterpart of magnitude $K_{\textrm s} = 18\fm0 \pm 0.1$ was identified by \citet{2009ATel.2233....1T} within the {\it Chandra} error box, but no radio counterpart was detected with a $3\sigma$ upper limit of 0.10 mJy \citep*{2009ATel.2232....1M}.
The source faded beyond \xte\ detection limit after 2009 October 11 (MJD 55113,  \citealt{2009ATel.2237....1M}).

Type I X-ray bursts were observed in \igr\ with {\it Swift} \citep{2009ATel.2198....1B} and burst oscillations immediately after with \xte\ (\citealt{2009ATel.2199....1W}, see \citealt{2010arXiv1012.0229F} for the analysis of all detected bursts). Several distance constraints have been reported based on these data.
The analysis of {\it Swift} data by \citet{2010A&A...509L...3B}  yielded an upper limit on the distance of $10.1 \pm 0.5$ kpc, derived using the empirical relation of the Eddington limit $L_{\rm edd} \approx (3.79 \pm 0.15) \times 10^{38}$ erg s$^{-1}$
for the photospheric radius expansion bursts \citep{2003A&A...399..663K}. Another upper limit of 7.5 kpc was derived by \citet{2010arXiv1012.0229F} via the independent analysis of type I bursts.
Using the same method, \citet{2010MNRAS.407.2575P} reported a similar upper limit as \citet{2010A&A...509L...3B} from {\it XMM-Newton} data, but the analysis of \xte\ data by \citet{2010MNRAS.tmp.1363A} gave a tighter upper limit of $6.9$ kpc.
\citet{2010MNRAS.tmp.1363A} also used the distance approximation of \citet{2008ApJS..179..360G}, that resulted in an upper limit of 4.4 kpc for a NS of mass $1.4 \msun$, radius $10$ km and hydrogen mass fraction $X = 0.7$.
The corresponding upper limit  for $X = 0$ would be 5.76 kpc, but absence of hydrogen is inconsistent with the fact that the companion
of \igr\ seems to be a main sequence star \citep{2010MNRAS.407.2575P, 2011A&A...526A..95R}.
In light of these constraints, we adopt the distance value of 5 kpc.

The source light curve  shows an exponential flux decay, commonly seen in AMPs (a ``slow decay" in the terminology of \citealt{2008ApJ...675.1468H}).
The pulse profiles are single-peaked, indicating that we probably see the emission coming mainly from  one emitting spot on the neutron star surface (i.e. contribution from the secondary spot  does not produce a distinct secondary feature; however, ``flattened'' pulse profile minima may suggest a  presence of secondary spot emission).

In this paper, we present the results of our spectral and timing analysis of \igr\ based on \textit{Swift} and \textit{RXTE} observations.
We study the evolution of phase averaged- and phase resolved spectra, phase lags and pulse profile changes during the outburst.

\section{Observational data}

The \textit{RXTE} data covering the outburst of \igr\ (ObsID 94041) were reduced using {\sc heasoft} 6.8 and the {\sc CALDB}.
We used data taken both by \textit{RXTE}/PCA (3--25 keV)  and HEXTE (25--200 keV).
Standard 0.5 per cent systematic was applied to the PCA spectra \citep{2006ApJS..163..401J}.
To keep the calibration uniform, we used data from PCA unit 2 only.
The source spectrum is contaminated by the Galactic ridge emission \citep{2009Natur.458.1142R}.
To take it into account for \xte/PCA spectra, we have produced a spectrum from the observations, where both \igr\ and nearby AMP XTE J1751--305 were in a quiescent state (MJD 55115 -- 55126).
This spectrum, mainly affecting channels below 15 keV, was subtracted from all spectral and timing data.

For \textit{Swift}/XRT observations, we only considered window-timing mode data, because the photon-counting mode data suffered from photon pile-up \citep{2010A&A...509L...3B}.
We reduced the \swift/XRT data with {\sc xrtpipeline} v.0.12.3 using standard filtering and screening criteria for the event selection.
We used circular regions of 20 pixel radius to extract the spectral data.
The {\sc xrtexpomap} task was used to generate the exposure maps and the ancillary response files were generated with the {\sc xrtmkarf} task to account for different extraction regions, vignetting and point spread function corrections.   Ancillary response files (ARFs) of  individual \swift/XRT snapshots were averaged together; each ARF was co-added with ``weight'' equal to the relative contribution of photons detected in the snapsot to the overall photon number collected from all snapshots. The \swift/XRT redistribution matrices (v.011) were taken from the {\sc CALDB}. \swift/XRT and \xte/HEXTE spectra were grouped such as each bin contained at least 200 counts.

The type I X-ray bursts \citep{2010A&A...509L...3B, 2010MNRAS.407.2575P, 2010MNRAS.tmp.1363A, 2011A&A...526A..95R, 2010arXiv1012.0229F}
were screened out from our analysis.
The spectral analysis was done using {\sc XSPEC} v.12 \citep{1996ASPC..101...17A}.
Uncertainties of spectral and timing best-fitting parameters correspond to the 90 per cent confidence level, unless otherwise stated.


\begin{table}
\caption{Data groupings}
\begin{tabular}{|lll|}
\hline
Group code &   MJD interval/\textit{RXTE} & MJD interval/\textit{SWIFT}  \\
\hline
T & 55087.9--55109.3 & 55087.8--55107.5 \\ 
1 & 55087.9--55088.9 & 55087.8--55088.7\\ 
2 & 55089.2--55090.5 & 55089.6--55090.8 \\ 
3 & 55091.2--55094.0 & 55092.8--55093.9 \\ 
4 & 55094.0--55096.8 & 55094.4--55095.5\\ 
5 & 55097.2--55099.5 & -- \\ 
6 & 55100.3--55101.9 & -- \\ 
7 & 55102.2--55104.9 & 55102.4--55105.0\\ 
8 & 55105.4--55109.3 & 55107.0--55107.5\\ 
\hline
\end{tabular}
\label{t:groups}
\end{table}

\section{Spectral analysis}
\label{spec_anal}
In this section, we describe the results of our spectral analysis of the source.
In order to improve statistics, we have grouped individual spectra as described in Table \ref{t:groups}.

\subsection{Spectral model}
\label{spec_model}
\begin{figure}
\centerline{\epsfig{file=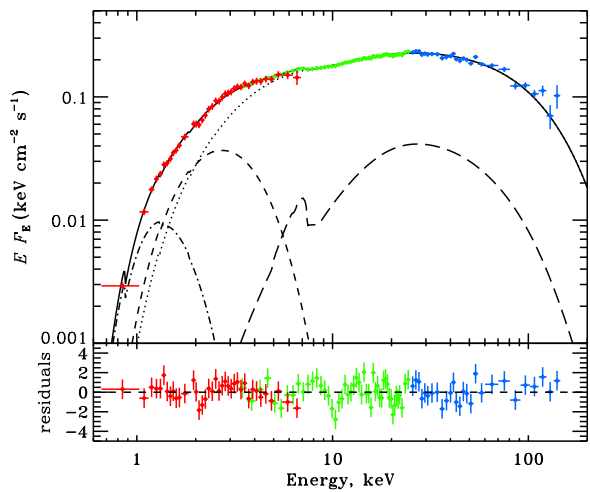,width=8cm}}
\caption{The joint observed spectrum of \igr\ from \textit{RXTE} and \textit{Swift} satellites collected during the entire outburst (group T). Red, green and blue data points represent \swift/XRT, \xte/PCA and \xte/HEXTE respectively. Solid, dotted, long-dashed, dashed and dot-dashed curves represent the total spectrum, Comptonization continuum, reflection and iron line, blackbody emission from the hotspot and the disc blackbody, respectively. Lower panel show the residuals of the fit. The fit parameters are given in Table \ref{t:fits} (group T for joint \xte\ and \swift\ spectra). Error bars correspond to 1$\sigma$.}
\label{f:spectrum}
\end{figure}

The spectrum of \igr\ is typical for an AMP and can be described as a composition of accretion disc emission (around 1--2 keV, not visible in the \textit{RXTE} range), blackbody originating from the hotspot (2--10 keV) and hard X-ray tail generated by thermal Comptonization in accretion shock located above the neutron star surface (contributing in the whole range of 1--200 keV and dominating above 10 keV), similar to other objects of this kind (e.g. \citealt{2002MNRAS.331..141G,2005MNRAS.359.1261G,2005A&A...436..647F,2005A&A...444...15F,2007A&A...464.1069F}; see Fig. \ref{f:spectrum}). To model thermal Comptonization continuum we used the  \textsc{compps} model  of  \citet{1996ApJ...470..249P}.
A fluorescent iron line at 6.4 keV and the Compton reflection of the \textsc{compps}  component \citep{1995MNRAS.273..837M} were also included in the fitting.  The spectral model includes also interstellar absorption model \textsc{phabs} parametrized by the hydrogen column density $N_\mathrm{H}$.
The described approach corresponds to \textsc{phabs*(diskbb+bbodyrad+ compps+diskline)} model in {\sc XSPEC}.

For \textsc{compps} we assumed a slab geometry and the model is characterized by the Thomson optical depth $\tau_{\rm T}$ and the   temperature of the hot electrons $\kte$.
The seed photons for Comptonization have a temperature $\ktseed$ and the surface area is denoted as $\Sigma_\mathrm{shock}$.
The blackbody component has a temperature \ktbb\ and its surface area is denoted as $\Sigma_\mathrm{spot}$.
The apparent spot radii at infinity can be computed from the \textsc{bbodyrad} and \textsc{compps} model normalizations: $\Sigma=\pi R^2= \pi K D^2_{10}$, where $K$ is the normalization obtained from fits, $R$ is the apparent radius in kilometres  and $D_{10}$ is the distance in units of 10 kpc.
We denote these radii as $R_\mathrm{spot}$ and $R_\mathrm{shock}$ for the \textsc{bbodyrad} and \textsc{compps}, respectively.
The Compton reflection from the accretion disc is parametrized by the  amplitude $\refl = \Omega / 2\pi$, where $\Omega$ is the solid angle covered by the reflecting medium \citep{1995MNRAS.273..837M}.
We used  \textsc{diskline} model  to model the iron line and we fixed the inner disc radius $r_{ \rm in} = 10 r_{\rm s}$, where $r_{\rm s} = 2GM / c^2$ is the Schwarzschild radius.
We also assumed that the radial emissivity profile for the illuminating continuum flux $\propto r^{-3}$.
For the \textsc{diskline}, as well as for \textsc{compps}, we assumed an inclination of $i = 60\degr$.
The \textsc{diskbb} model component was only included in the joint analysis of \swift\ and \xte\ data (Section \ref{s:swift}) to account for the soft X-ray accretion disc emission. The respective model parameters are the inner disc temperature \tin\ and the apparent inner disc radius $\rin$, which can be computed from the model normalization $K_\mathrm{dbb}$ as $\rin = D_{10} \sqrt{K_\mathrm{dbb} / \cos i}$.

\begin{table*}
\centering
 \begin{minipage}{150mm}
\caption{Results of spectral fitting with the Comptonization model with constant areas for
the blackbody and Comptonization components (Sect. \ref{s:swift},  \ref{c:rxte}).
Left column indicates the data group. Letter ``F" indicates a fixed parameter.\label{t:fits}   }
\begin{tabular}{|lllllllllll|}
\hline
Group   &  \kte     & $\tau_{\rm T}$  &  \ktseed  &  $\refl$  & \ktbb  &  $\Sigma$ & $EW$   &  \ktdisk  & $\rin$  & $\chi^2$/d.o.f \\
  &  keV     &  &  keV  &  & keV  &   km$^2$ &  eV   &   keV  & km &   \\
\hline
T \footnote{\xte\ and \swift, interstellar absorption yield $\nh=(0.88_{-0.24}^{+0.21}) \times 10^{22} \mathrm{cm}^{-2}$.}
&  $30 \pm 2$         & $1.80 _{-0.10}^{+0.12 }$     &$ 1.05 _{- 0.11 }^{+ 0.10 }$ &  $0.34\pm0.09$               & $0.62 \pm 0.07 $               & $49 _{- 19 }^{+ 25 }$ & $ 62 _{- 24 }^{+ 25 }$   & $0.24\pm 0.07$ & $40_{-27}^{+46} $ & 271/347 \\
T \footnote{\xte\ only.}   & $ 31 \pm 2 $        &  $ 1.77 _{- 0.08 }^{+ 0.11 }$&$ 1.00 _{- 0.17 }^{+ 0.11 }$ &  $ 0.36 \pm 0.08 $           &  $ 0.58 _{- 0.11 }^{+ 0.08 }$  & $ 59 _{- 19 }^{+ 64 }$& $ 57 _{- 23 }^{+ 25 }$   & & & 193/161  \\
1--4    &$ 33 _{- 3 }^{+ 2 }$ & $ 1.70_{- 0.08 }^{+ 0.12 }$  &$ 1.01 _{- 0.17 }^{+ 0.11 }$ &  $ 0.40 _{- 0.10 }^{+ 0.09 }$&   $ 0.59 _{- 0.12 }^{+ 0.08 }$ & $ 93 _{- 23 }^{+ 77 }$ & $ 73 _{- 23 }^{+ 25 }$ & & & 151/161\\
5--8    & $ 28 \pm 2 $        &  $ 1.88 _{- 0.10}^{+ 0.13 }$ &$ 1.00 _{- 0.20 }^{+ 0.13 }$ &  $ 0.33 _{- 0.11 }^{+ 0.12 }$&   $ 0.60 _{- 0.13 }^{+ 0.09 }$ & $ 46 _{- 17 }^{+ 65 }$ &  $ 44 _{- 26 }^{+ 30 }$  & & & 208/161 \\
\hline
1  &  30F&  $ 1.85 _{- 0.02 }^{+ 0.01 }$    &  $ 1.09 _{- 0.10}^{+ 0.08 }$  & 0.3F &    $ 0.66 _{- 0.07 }^{+ 0.05 }$    &   $ 63 _{- 14 }^{+ 31 }$ & $ 95 \pm 22 $              &   &   &   139/163  \\
2  &  30F&  $ 1.86 _{- 0.01 }^{+ 0.02 }$    &  $ 1.16 _{- 0.07 }^{+ 0.06 }$ & 0.3F &    $ 0.71 _{- 0.05 }^{+ 0.04 }$    &   $ 48 _{- 8 }^{+ 13 }$  & $ 106 _{- 20 }^{+ 19 }$    &   &   &   163/163  \\
3  &  30F&  $ 1.86 _{- 0.02 }^{+ 0.01 }$    &  $ 1.06 _{- 0.10}^{+ 0.07 }$  & 0.3F &     $ 0.64 _{- 0.07 }^{+ 0.05 }$   &   $ 61 _{- 14 }^{+ 29 }$ &  $ 102 \pm 20 $            &   &   &   149/163  \\
4  &  30F&  $ 1.82 \pm 0.02 $               &  $ 1.09 _{- 0.13 }^{+ 0.08 }$ & 0.3F &    $ 0.65 _{- 0.09 }^{+ 0.05 }$    &   $ 46 _{-11}^{+29}$  & $ 76 \pm 21 $              &   &   &   163/163  \\
5  &  30F&  $ 1.84 _{- 0.01 }^{+ 0.02 }$    &  $ 1.11 _{- 0.13 }^{+ 0.07 }$ & 0.3F &     $ 0.66 _{- 0.09 }^{+ 0.05 }$   &   $ 38 _{-9 }^{+22}$  &  $ 72 \pm 20 $             &   &   &   167/163  \\
6  &  30F&  $ 1.85 _{- 0.03 }^{+ 0.08 }$    &  $ 1.02 _{- 0.45 }^{+ 0.12 }$ & 0.3F &     $ 0.60_{- 0.09 }^{+0.07 }$    &     $46_{-16}^{+40}$   &  $ 53 \pm 23 $             &   &   &   171/163  \\
7  &  30F&  $ 1.80 \pm 0.03 $               &  $ 1.01 _{- 0.14 }^{+ 0.11 }$ & 0.3F &     $ 0.57 _{- 0.09 }^{+0.08 }$   &   $ 39 _{-13 }^{+30 }$ &  $ <39 $      &   &   &   213/163  \\
8  &  30F&  $ 1.74 \pm 0.03 $               &  $ 0.98 _{- 0.15 }^{+ 0.09 }$ & 0.3F &     $ 0.56 _{- 0.10}^{+0.06 }$    &   $ 37 _{-12 }^{+32 }$ &  $ <30 $         &   &   &   175/163  \\
\hline
\end{tabular}
\end{minipage}
\end{table*}


\begin{table*}
\caption{Results of spectral fitting with the Comptonization model using independent areas of the hotspot blackbody and Comptonized component using \xte\ data only, Sect. \ref{c:rxte}. Left column indicates the data group. \label{t:diffareas} }
\begin{tabular}{|llllllllll|}
\hline
Group   &  \kte    & $\tau_{\rm T}$  &  \ktseed  &  $\refl$  & \ktbb  & $EW$  &  $\Sigma_\mathrm{spot}$ & $\Sigma_\mathrm{shock}$ & $\chi^2$/d.o.f \\
    &   keV     &   &   keV  &   &  keV  & eV   &   km$^2$ &  km$^2$ &   \\
\hline
T       &    $ 33 _{- 3 }^{+ 4 }$     &      $ 1.61 \pm 0.18 $         &    $ 1.15 _{- 0.21 }^{+ 0.30}$     &   $ 0.45 _{- 0.12 }^{+ 0.13 }$   &   $ 0.60 \pm 0.13 $             &  $ 46 _{- 31 }^{+ 34 }$& $ 63 _{- 26 }^{+ 124 }$  &  $ 34 _{- 21 }^{+ 42 }$  &  190/160  \\
1--4    &    $ 35 \pm 3 $             &   $ 1.54 _{- 0.15 }^{+ 0.18 }$ &    $ 1.16 _{- 0.19 }^{+ 0.25 }$     &   $ 0.49 _{- 0.13 }^{+ 0.14 }$   &   $ 0.61 _{- 0.13 }^{+ 0.11 }$ &  $ 61 _{- 31 }^{+ 33 }$   &  $ 80 _{- 31 }^{+ 134 }$  &  $ 43 _{- 24 }^{+ 48 }$     &  148/160  \\
5--8    &    $ 31 _{- 3 }^{+ 6 }$     &   $ 1.66 _{- 0.29 }^{+ 0.27 }$ &    $ 1.21 _{- 0.29 }^{+ 0.34 }$    &   $ 0.46 _{- 0.18 }^{+ 0.22 }$   &   $ 0.63 _{- 0.15 }^{+ 0.11 }$ &  $ <110          $       &  $ 44 _{- 17 }^{+ 88 }$    &  $ 23 _{- 14 }^{+ 42 }$     &   206/160 \\
\hline
\end{tabular}
\end{table*}

\subsection{Phase-averaged spectra from \textit{RXTE} and \textit{Swift}}
\label{s:swift}

\begin{figure}
\centerline{\epsfig{file=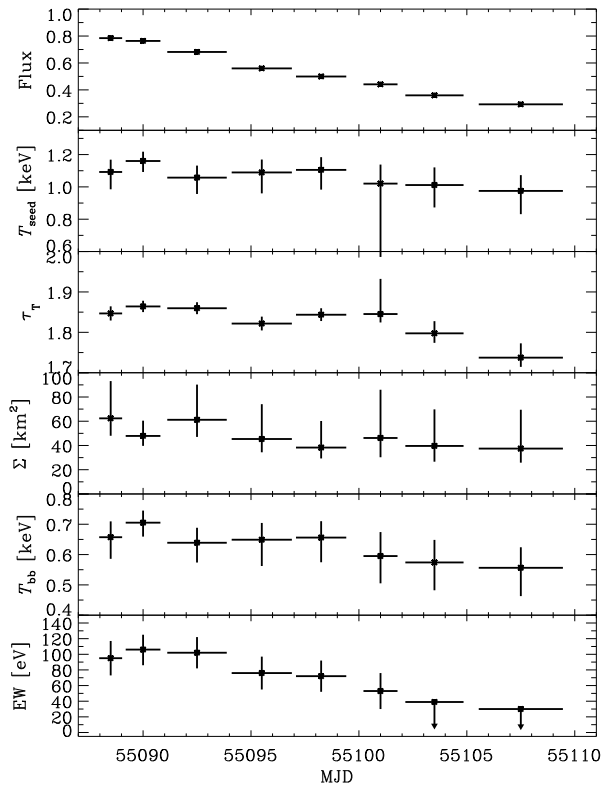,width=8cm}}
\caption{The best-fitting parameters for the Comptonization-based model of Section \ref{c:rxte}. The absorption corrected flux in 3--20 keV band is in units of $10^{-9}$ erg cm$^{-2}$ s$^{-1}$.
}
\label{f:compps}
\end{figure}

\begin{figure*}
\centerline{\epsfig{file=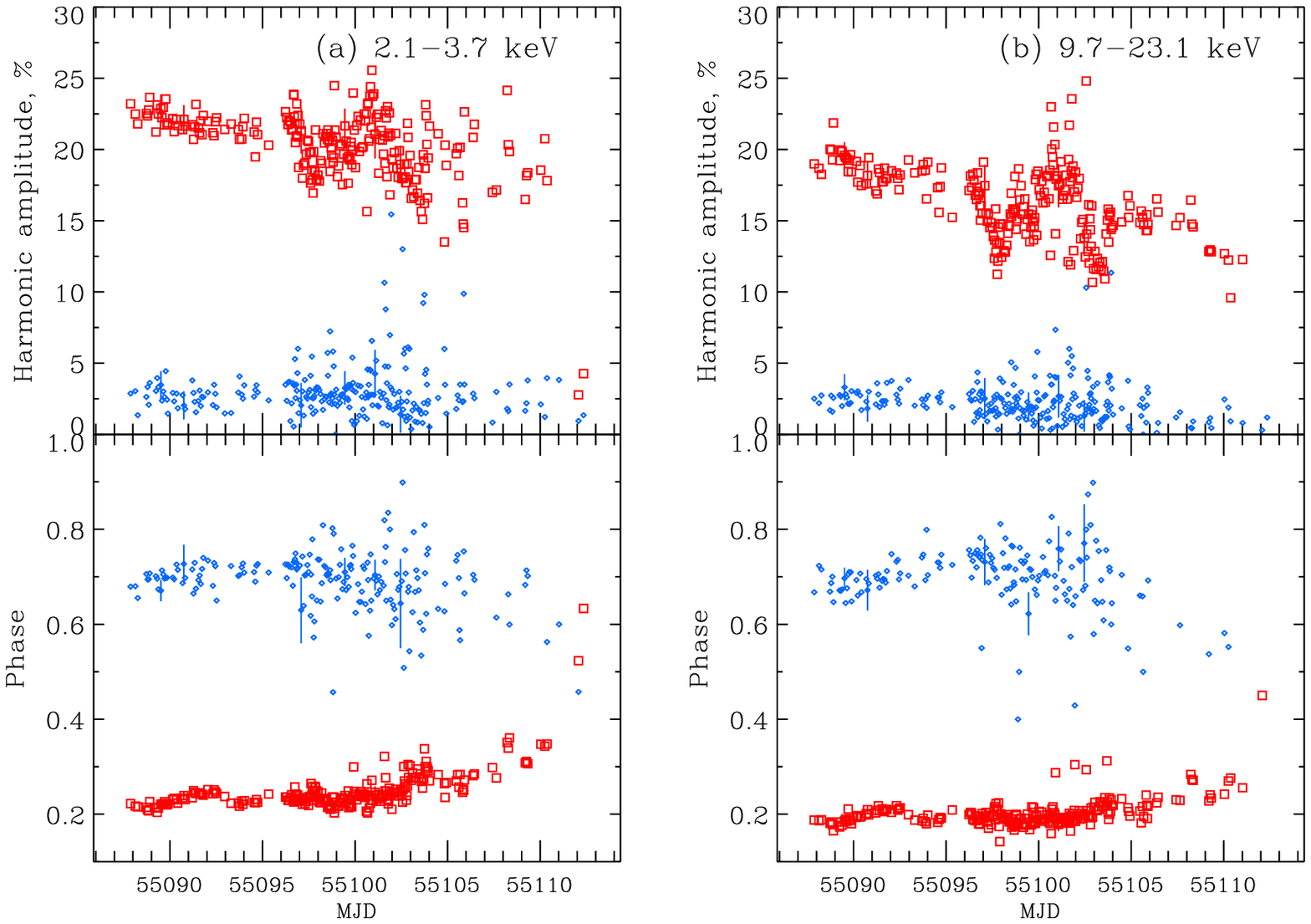,width=16cm}}
\caption{Results of fitting the per-orbit pulse profiles in (a) 2.1--3.7 keV  and (b) 9.7--23.1 keV with expression (\ref{m:cosines}). Top panels: the amplitudes of the  fundamental $a_1$ (squares, red) and overtone $a_2$ (diamonds, blue). Bottom panels: phases of the fundamental $\phi_1$ (squares, red) and overtone $\phi_2$ (diamonds, blue, shifted by phase 0.5 for clarity). Typical error bars are shown for a few points. Note the changes in amplitude and  phase of the fundamental  around MJD 55112, addressed in Section \ref{c:mjd55112}. }
\label{f:sinefits}
\end{figure*}

We used the \textit{Swift}/XRT data (in range 0.6--8.0 keV) to study the soft X-ray emission of \igr.
Initially, we fitted group T to constrain the disc parameters and the absorption column. The emitting areas of the hotspot and the shock were assumed equal (while the spectrum is statistically good to fit these areas independently, a large number of free parameters would lead to large uncertainties; in fact, these areas must be similar on physical grounds, see \citealt{2009MNRAS.400..492I}). The best-fitting results are shown in Table \ref{t:fits}; in particular we find that the emission below $\lesssim 3$ keV has clear signatures of the accretion disc \citep[see Fig. \ref{f:spectrum} and also][fig. 9]{2010MNRAS.407.2575P}.
Fits with  free interstellar absorption lead to $\nh=(0.88_{-0.24}^{+0.21} )\times 10^{22} \mathrm{cm}^{-2}$, $\ktdisk =0.24\pm0.07$ keV and $\rin=40_{-27}^{+46}$ km.

The aforementioned best fitting values are subjected to several uncertainties. The values of the disc parameters \tin\ and \rin\ are tightly correlated with the absorption value $N_\mathrm{H}$. In addition, the derived value of \rin\ depends on the assumed distance and it should be corrected for two effects in order to obtain a realistic radius. As discussed in \citet{1998PASJ...50..667K} and \citet{1999MNRAS.309..496G}, the derived radius should be larger by the square of the colour correction factor $f_{\rm c}$.  This factor $f_{\rm c} = 1.7$ was computed for accretion discs around black holes by \citet{1995ApJ...445..780S}.
However, in the case of AMPs, there is a stark difference in the sense that the disc is irradiated by the emission from the hotspot, which casts an uncertainty to this value.
Furthermore, \rin\ is also affected by a correction due to the inner boundary condition \citep{1999MNRAS.309..496G},
but this (of the order of unity) factor is not accurately known in the case of accretion onto a magnetized star.
Therefore, the value of the inner disc radius should be taken as an order of magnitude estimate.

Other spectral parameters (see Table \ref{t:fits}) are consistent with the findings of \citet{2010arXiv1012.0229F}.
However, there are small differences between our results and \citet{2010MNRAS.407.2575P} especially for the values of $\tau_{\rm T}$ and $\ktseed$.
The most likely reason for these differences is because the high energy cutoff cannot be accurately determined in short HEXTE exposures.
Also, the cross-calibration issues between \swift/XRT and \xmm/EPIC instruments \citep[see][]{2011A&A...525A..25T} might cause the differences.

In further analysis, we found that the \textit{Swift}/XRT data do not have enough statistics to reliably constrain the disc component for individual fits of groups $1$--$8$.
Therefore, we could not look for changes in the disc parameters during the outburst and in the following sections we only consider \xte\ data when we study the evolution of the spectral parameters.

\subsection{Phase-averaged spectra from \textit{RXTE}}
\label{c:rxte}

We began our $\xte$--only spectral analysis by fitting the model with independent emitting areas for the hotspot and the shock (free blackbody and Comptonized component normalizations). Because the accretion disc does not contribute to the flux in the PCA band 3--25 keV, we omit the {\sc diskbb} component from the following spectral fits and use our best fitting value of $\nh=0.88\times 10^{22} \mathrm{cm}^{-2}$ in the fitting of \xte-only data.
The spectrum for group T and ``joined" groups 1--4 and 5--8 have sufficient statistics to fit these areas separately.

The results of the fitting are shown in Table \ref{t:diffareas}.
We find that the ratio between the areas is compatible with constant, although the statistical errors are large to make a firm conclusion. We note that analysis of the same source by \citet{2010MNRAS.407.2575P} and of SAX J1808.4--3658 by \citet{2009MNRAS.400..492I} both suggest, in agreement with our result, that the blackbody  area is 2--3 times larger than the area of Comptonized emission. These fits indicate a decrease of the emitting areas, as expected from a gradually increasing inner disc radius \citep*{2009ApJ...706L.129P}.
The simultaneous decrease of the iron line equivalent width supports the expected physical picture (note that the reflection amplitude should decrease as well, but observation uncertainties do not allow us to constrain it reliably).

The group T and the ``joined" groups 1--4 and 5--8 also allow for independent fitting of \kte\ and reflection amplitude $\refl$.
The best fitting parameters are shown in Table \ref{t:fits}. We note that the actual value of $\refl$ is subject to the chosen continuum model (e.g., \citealt*{2007A&ARv..15....1D}).

Spectra for  individual groups $1$--$8$ do not have enough statistics to constrain \kte\ and $\refl$ (that turns out to be uncertain and compatible with zero).
Therefore, we fixed \kte=30 keV and $\refl=0.3$ (as found in the \textit{Swift} and \textit{RXTE} fits, Sect. \ref{s:swift}) and obtained fits for individual data groups.
Furthermore, we adopted equal emitting areas for the Comptonization and blackbody components for these groups, because the data quality does not allow us to fit them independently and they should be similar on physical grounds (see \citealt{2005MNRAS.359.1261G, 2009MNRAS.400..492I}).

The time evolution of the spectral parameters is shown in Fig. \ref{f:compps} and in Table \ref{t:fits}.
The optical depth is initially roughly constant $\tau_{\rm T} \approx 1.85$ for groups $1$--$7$, but later it drops to $\tau_{\rm T} = 1.74\pm0.03$ for group $8$.
This shows (in connection with fixed \kte\ value) that the spectrum softens in the end of the outburst.
It is also noticeable, that $\Sigma_\mathrm{spot}$, \ktseed\ and \ktbb\ are decreasing slightly during the outburst.
The decrease in $\Sigma_\mathrm{spot}$ is most likely caused by the change of $\rin$ in the course of the outburst \citep{2009ApJ...706L.129P}.
When the flux drops as the mass accretion rate goes down, $\rin$ increases (it is likely proportional to the Alfv{\'e}n radius which has a $\dot{M}^{-2/7}$ dependence, see e.g. \citealt*{FKR02}).
Assuming that the magnetic field is a dipole, the outer boundary of the hotspot is controlled by the current position of $\rin$
and therefore the increase in $\rin$ leads to decrease in $\Sigma_\mathrm{spot}$.

\section{Timing analysis}

The pulse shape of an accreting pulsar contains important information about physics of emission and geometrical parameters of the system (\citealt{2003MNRAS.343.1301P}, \citealt*{2008ApJ...672.1119L}; \citealt{2008AIPC.1068...77P, 2009MNRAS.400..492I, 2009ApJ...706L.129P}).
To obtain the pulse profiles, we used the ephemeris of \citet{2009ATel.2221....1R}.
In general, the pulse profiles of \igr\ are single-peaked, close to symmetric and without a prominent secondary maxima.

\begin{figure*}
\centerline{\epsfig{file=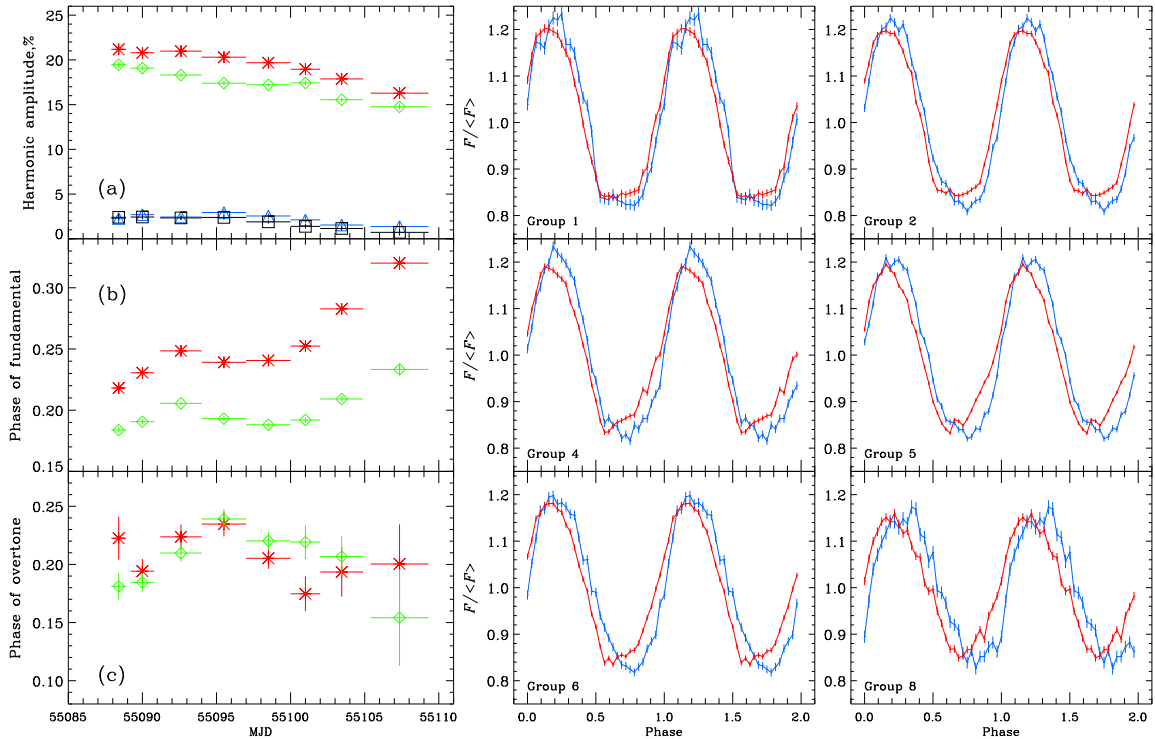,width=16cm}}
\caption{Left panels: evolution of harmonics content for groups 1--8. Stars and triangles represent $2.1$--$3.7$ keV, diamonds and squares -- $9.7$--$23.1$ keV. (a)  The amplitudes of the  fundamental $a_1$  (upper points) and overtone $a_2$ (lower points); (b) phases of the fundamental $\phi_1$  and (c) overtone $\phi_2$ versus time. Right panels: pulse profiles for $2.1$--$3.7$  and $9.7$--$23.1$ keV (blue and red histograms, respectively). For amplitudes and phases (panels a--c) the errors are
at the  90 per cent confidence level, while for pulse profiles the errors bars  correspond to $1\sigma$.}
\label{f:pulses}
\end{figure*}

\begin{figure*}
\centerline{\epsfig{file=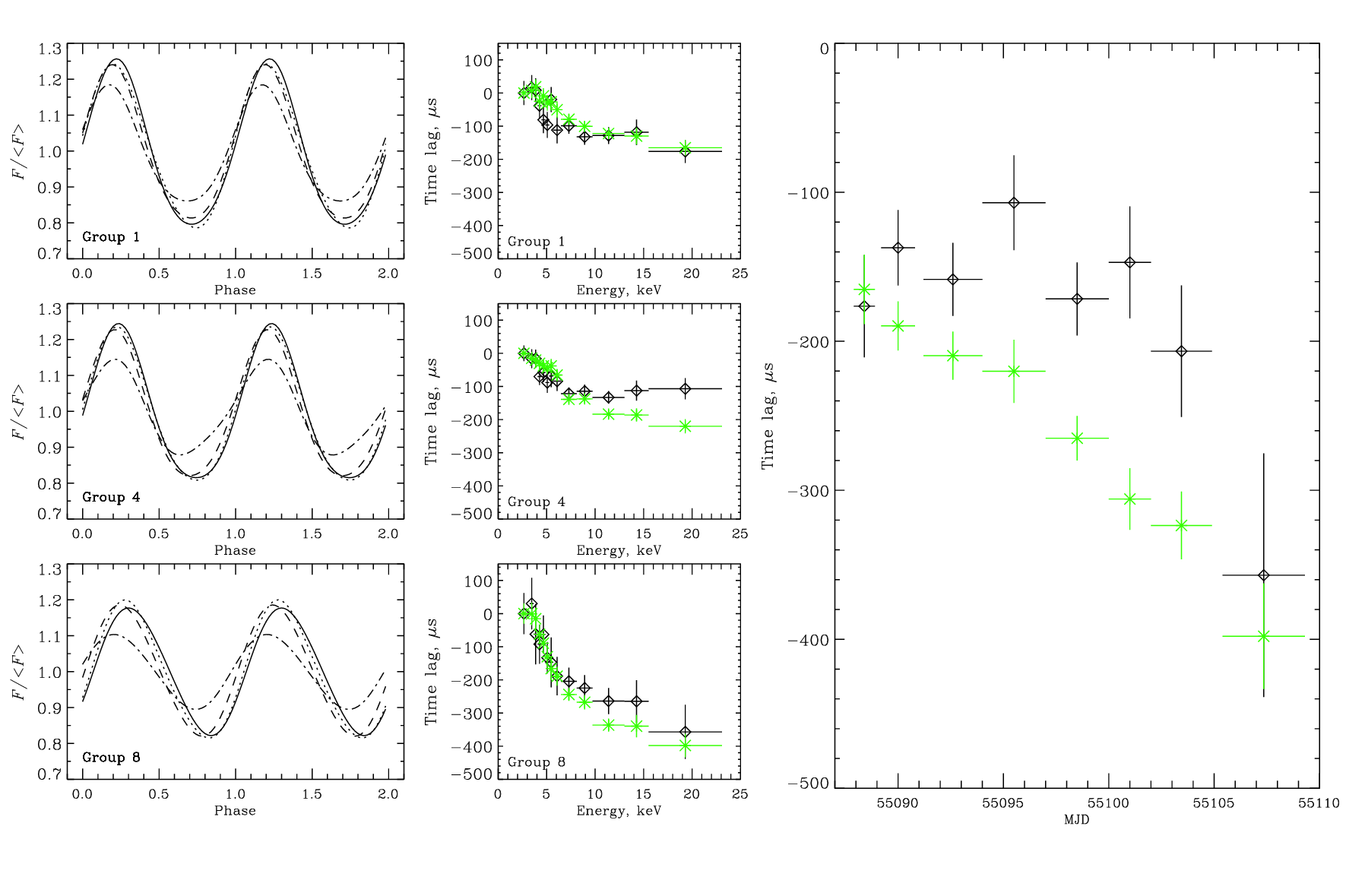,width=16cm}}
\caption{Left panels: pulse profile fits with equation (\ref{m:cosines}) for energies 3.3--3.7, 4.5--4.9, 6.5--8.1 and 15.5--23.1 keV (solid, dotted, dashed and dash-dotted lines, respectively). Middle panels: pulse maximum lags (stars) and the lags of the fundamental (diamonds).
The overtone lags have large errors and therefore are not shown here. The observational groups are indicated on the panels.
Right panel: evolution of the lags between 2.1--3.7 and 15.5--23.1 keV energy bands of the pulse maximum and of the fundamental.
The pulse maximum lag was computed as the phase difference between the maxima at different energies,
determined via fitting the observed pulse shape with expression (\ref{m:cosines}). }
\label{f:lags}
\end{figure*}

\subsection{Harmonic content evolution}

From our spectral analysis (e.g., Fig. \ref{f:spectrum}) we see that the hard X-ray part of the spectrum above 10 keV is dominated by thermal Comptonization, that probably takes place in the accretion shock.
The soft part (below 10 keV) includes blackbody emission from the hotspot.
Furthermore, around 2 keV there is emission from the accretion disc \citep[see Fig. \ref{f:spectrum} and][fig. 9]{2010MNRAS.407.2575P}.
Consequently, we chose energy ranges of interest as $2.1$--$3.7$ and $9.7$--$23.1$ keV: the first interval contains a large fraction of the hotspot blackbody radiation (and some part of  disc emission which is non-pulsating), while the second band contains only Comptonized emission. While a throughout analysis by \citet{2011A&A...526A..95R} demonstrated that one needs three (and, sometimes, even four) harmonics to fully describe wide-energy (2--25 keV) pulses,  we find it acceptable to utilize two Fourier harmonics for description of pulses in narrow energy bands and trace the pulse profile changes. We note that wide-energy pulse profile is in fact a superposition of different pulse shapes seen at different energies and this might affect the harmonics decomposition.
However, indeed the comparison between two-harmonics fit and the actual pulse profiles reveal some  deviations from a smooth fit near the  pulse profile minima, that can be due to the antipodal spot contribution  or due to additional absorption at certain phase (by e.g. the accretion column), which  \citet{2011A&A...526A..95R} modelled with a  3rd harmonic (as it follows from their fig. 2). We fitted the pulse profiles collected from each \textit{RXTE} orbit with the following expression
\be
\label{m:cosines}
F(\phi)=\overline{F}\{ 1+a_1\cos[2\pi(\phi-\phi_1)] +a_2\cos[4\pi(\phi-\phi_2)] \} ,
\ee
where $a_1$, $a_2$ are amplitudes and $\phi_1, \phi_2$ are phases of the fundamental and the overtone, respectively.

The fitting results for  aforementioned energy bands are shown in Fig. \ref{f:sinefits}.
The amplitude of the fundamental decreases with time (and with flux).
While the general trend is rather smooth, some irregularity in fundamental amplitude can be noticed around MJD 55098, and phase of overtone experiences small shift around MJD 55093.
After MJD 55100, we observe a ``drift" in the phase of fundamental, seen clearly on Fig. \ref{f:pulses} (see also fig. 4 in \citealt{2011A&A...526A..95R}).
This drift is larger at soft energies, which in turn affects the value of the phase lag (see Sect. \ref{s:obslags}).
The harmonic content on soft and hard energies behaves very similarly, which is also seen in SAX J1808.4--3658 during its ``slow decay" stage \citep{2009MNRAS.400..492I}.
After the slow decay, SAX J1808.4--3658 has shown quite a different evolution of pulse shapes at different energies. For our case, there is a clear evidence that the source experienced a likely transition to the ``rapid drop'' stage (see Section \ref{c:mjd55112}).
However, its flux dropped very quickly below the detection level, making it impossible to study pulse profile evolution further.

\subsection{Evolution of phase lags}
\label{s:obslags}

In AMPs the pulses in the soft and hard energies do not arrive in phase, but there is an energy-dependent phase lag (e.g., \citealt{1998ApJ...504L..27C}).
We determined the  phase lags by fitting the pulse profiles at a given energy with expression (\ref{m:cosines}) and finding the phase difference relative to the reference energy band 2.1--3.7 keV. This way we find the phase lags in each harmonic as well as
the pulse maximum lag corresponding to the phase difference between fitted pulse maxima.
In \igr, the phase lag is negative (i.e. pulse peaks on soft energies at a later phase) and shows a gradual increase from 3 keV to approximately 10 keV, where the lag value nearly ``saturates". This behaviour is typical for AMPs (see e.g.  \citealt{1998ApJ...504L..27C, 2002MNRAS.331..141G}, \citealt*{2009ApJ...697.2102H}, \citealt{2010arXiv1012.0229F}).
The phase lag of the overtone is poorly constrained: for our groups 1--8  it is noticeable that overtone best-fitting value decreases from 0 to $\sim-100$ $\mu$s in the energy range 3--7 keV, and then remains constant or slightly reduces. But in all cases it is compatible with zero and is determined with the uncertainty more than $100\mu$s.

\begin{figure}
\centerline{\epsfig{file=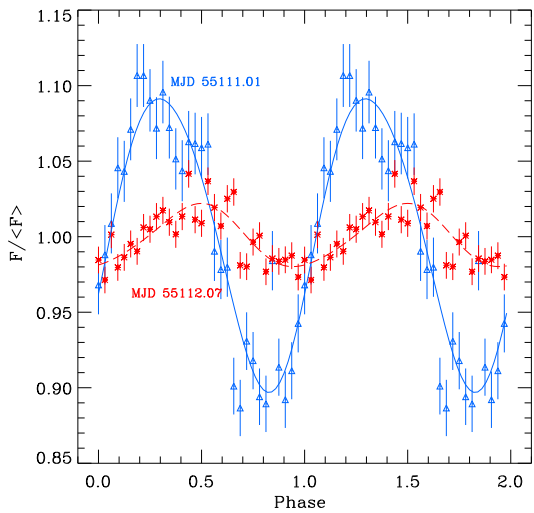,width=7cm}}
\caption{The pulse profiles (2--60 keV) from MJD 55111.01 (blue triangles) and MJD 55112.07 (red asterisks), demonstrating a significant shift of the pulse maximum and an abrupt drop in the amplitude (due to coincidental outburst of  a neighbouring  AMP XTE J1751--305).
Blue solid and red dashed curves are the respective best-fitting approximations with expression (\ref{m:cosines}). Error bars
correspond to 1$\sigma$.}
\label{f:mjd55112}
\end{figure}

The most noticeable effect we saw in the data is a gradual increase of the phase lag (measured between 2.1--3.7 and 15.5--23.1 keV) from 200 to 400 $\mu$s during the outburst, as illustrated on the right panel of Fig. \ref{f:lags}.
Interestingly, fundamental shows a straightforward trend, while pulse maximum lag changes noticeably only close to the end of the outburst.
A comparison of the pulse profiles obtained at various dates indicates that the lag increases because the pulse maximum at soft energies (2.1--3.7 keV in our case) shifts to a latter phase, while the maximum of the high-energy pulse (9.7--23.1 keV) shifts in parallel, but in a less pronounced way.
To illustrate that, in Fig. \ref{f:pulses} we plot a set of pulse profiles from a few time intervals and the evolution of their  harmonic content.

\subsection{Pulse profile changes at MJD 55112}


\begin{figure}
\centerline{\epsfig{file=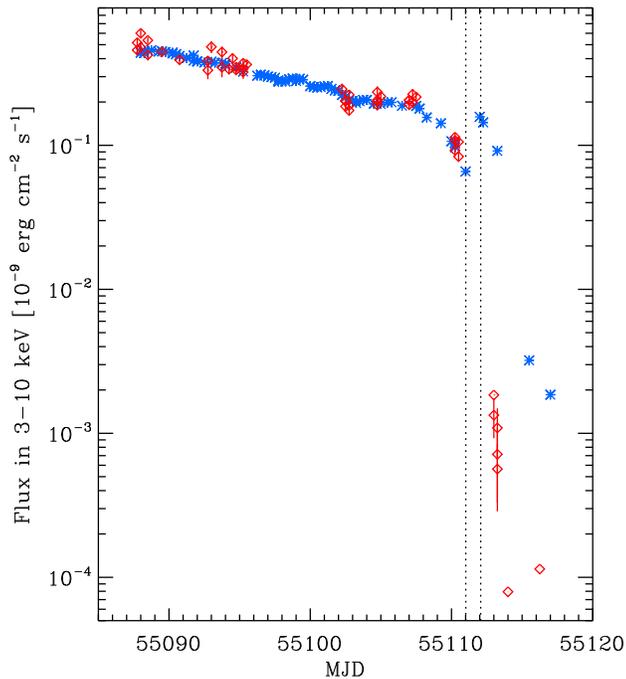,width=8cm}}
\caption{The lightcurve of \igr\ during the 2009 outburst.
The flux in the 3--10 keV range is corrected for absorption.
Blue stars correspond to the  \xte/PCA data, and red diamonds are for \swift/XRT (Photon Counting mode).
Two vertical dotted lines denote observations  centered at MJD 55511.01 and MJD 55112.07 that are shown in Fig. \ref{f:mjd55112}.
Although the \xte/PCA observations are contaminated   after MJD 55112 by the outburst of a nearby AMP XTE J1751--305,
\swift\ allows to determine the flux of  \igr\ unambiguously revealing an abrupt flux fall consistent with being the start of the
``rapid drop'' outburst stage. Error bars correspond to 1$\sigma$. }
\label{f:lc}
\end{figure}
\label{c:mjd55112}

The pulsations became completely undetectable shortly after MJD 55112, and the last useful observations do not contain a reliable statistics to obtain a well-defined energy-resolved pulse shape.
However, the shape in the whole \textit{RXTE}/PCA range (approximately 2--60 keV) can be used to locate the pulse maximum.
In Fig. \ref{f:mjd55112} we show two pulse profiles from the adjacent observations centered at MJD 55111.01 and MJD 55112.07 together with the respective harmonical fits. The relative amplitude abruptly decreases because the average X-ray emission is modified by simultaneous outburst of AMP XTE J1751--305 in the field of view of \textit{RXTE}/PCA \citep{2009ATel.2237....1M}.
It is clear that the phase of the pulse maximum has shifted forward by about 0.1--0.2. By analogue with SAX J1808.4--3658, we can speculate that the source began to shift into the rapid drop stage \citep{2009MNRAS.400..492I}.
As noted by \citet{2010arXiv1012.0229F}, the exponential trend typical to the slow decay outburst stage is followed by a faster, linear drop of flux around MJD 55107, few days prior to detected pulse evolution. Fig. \ref{f:sinefits} reveals that in parallel with the linearly
decreasing flux trend, the phase of fundamental starts to increase slowly, ending up in a sharp phase jump at MJD 55112. We can speculate that the accretion disc starts to recede from the neutron star and at some point the antipodal spot appears to our view  changing the pulse shape.

While \xte/PCA is a non-imaging instrument and it is impossible to separate flux from \igr\ and XTE J1751--305, we were able to
estimate flux received from \igr\ using \swift/XRT.
The \swift\ count rate lightcurve has been produced using online XRT product generator\footnote{http://www.swift.ac.uk/user\_objects/} \citep{2009MNRAS.397.1177E} and converted to energy flux in 3--10 keV band using the webPIMMS tool\footnote{http://heasarc.nasa.gov/Tools/w3pimms.html}
 using powerlaw with photon index 1.7 that is suitable for our object in the mentioned energy interval.
We also obtained the \xte/PCA light curve in the same energy range for cross-calibration.
The resulting lightcurve of the outburst is shown on Fig. \ref{f:lc}.
A sharp abrupt fall of the flux  has been observed on October 9 (MJD 55113,  \citealt{2009ATel.2237....1M}) followed by a non-detection of the source in the subsequent pointings. This strongly supports the conclusion that the source entered the rapid drop stage.
Since there is no \swift\ observation around MJD 55112, it is not possible to reliably correct the amplitude of the respective pulse profile shown on Fig. \ref{f:mjd55112}.

\subsection{Phase-resolved spectrum}

\label{c:phares}
\begin{figure*}
\centerline{\epsfig{file=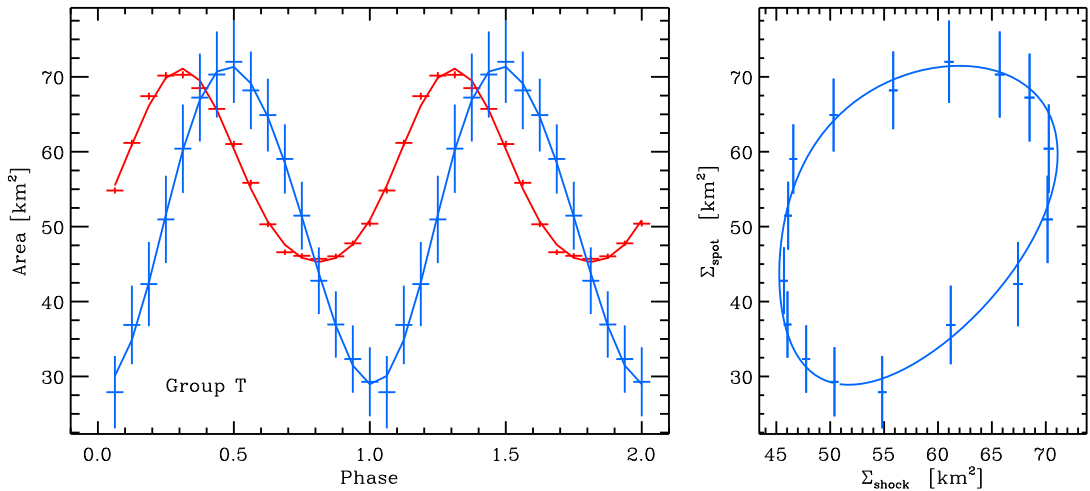,width=15cm}}
\caption{Results of the phase-resolved spectral analysis. The best-fitting parameters (except normalizations) are frozen at the values obtained in the corresponding phase-averaged fit (Sect. \ref{c:rxte}, Table \ref{t:fits}). Left panel: apparent emitting areas of the two components (blue and red curves represent blackbody $\Sigma_\mathrm{spot} $  and Comptonization $\Sigma_\mathrm{shock}$ components, respectively). Right panels: $\Sigma_\mathrm{spot} $  vs $\Sigma_\mathrm{shock}$. The parameters of the fits with a harmonic function are shown in Table \ref{t:normfits}. }
\label{f:phares}
\end{figure*}

Phase-resolved spectroscopy was performed for 1998 outburst of SAX J1808.4--3658 by \citet{2002MNRAS.331..141G}, for the 2002 outburst by \citet{2009MNRAS.400..492I} and \citet{2010MNRAS.tmp.1560W} and for XTE J1751--305 by \citet{2005MNRAS.359.1261G}.
In the former work it was concluded that the energy dependence of the pulse profiles and phase lags can be explained by a simple model where only normalizations of the hotspot blackbody and Comptonization tail vary.
Similar to these papers, we have generated the phase-resolved spectra for group T and used the phase-averaged spectrum (Section \ref{c:rxte}) as a reference.
We fixed all the parameters except the blackbody and Comptonization normalizations.
The results are shown in Fig. \ref{f:phares}.
The coefficients for expression (\ref{m:cosines}) that describe the modulation in the apparent areas of the blackbody $\Sigma_\mathrm{spot}$ and Comptonized tail $\Sigma_\mathrm{shock}$ are shown in Table \ref{t:normfits}.
Lag between components is large: fundamental components have phase difference of 0.18, which corresponds to $\sim$730$\mu$s, while the lag of the overtone is difficult to determine precisely. Phase-resolved fits of groups 1--4 and 5--8 suggest slight increase of component lags that reflects the changes in  the pulse profile. The determined effective areas are shifted comparing to Table \ref{t:diffareas} values, since we use continuum shape determined using  equal areas of blackbody and Comptonized emission (i.e., given in Table \ref{t:fits}). Utilizing continuum parameters obtained with independent emitting areas yield in area values compatible with those in Table \ref{t:diffareas}; however, the observed phase difference remains the same.

It is interesting to note that the measured  $\sim$730$\mu$s lag between components is much larger than the one observed in actual pulsations. The effect also seen in similar studies, e.g., of SAX J1808.4--3658 \citep{2002MNRAS.331..141G, 2009MNRAS.400..492I}. It is a natural consequence of the fact that in phase-resolved spectroscopy we decompose the observed joint spectrum into separate physical components, while the pulse that we observe in the soft X-ray band is a mix from these components (in nearly equal proportion, see Fig. \ref{f:spectrum}). Modelling of time lags (described below in Section \ref{s:lags}) confirms the fact  that the lag between components is indeed much larger than the one between the observed pulses at different energies.




\begin{table}
\caption{Harmonic fits by expression (\ref{m:cosines}) to the phase-resolved apparent areas.
 The average apparent areas $\overline{\Sigma}$ (in units of km$^2$)  are computed for the distance of 5 kpc.}
\begin{tabular}{|llllllll|}
\hline
Group		& Model component  &  $\overline{\Sigma}$ & $a_1$ & $a_2$ & $\phi_1$ & $\phi_2$ \\
\hline
T    & Spot  & 51 & 0.42 & 0.02 & 0.50 & 0.33\\
T    & Shock & 57 & 0.23 & 0.03 & 0.31 & 0.31 \\
1--4 & Spot  & 63 & 0.43 & 0.04 & 0.48 & 0.40 \\
1--4 & Shock & 70 & 0.23 & 0.03 & 0.31 & 0.30 \\
5--8 & Spot  & 40 & 0.45 & 0.03 & 0.52 & 0.27  \\
5--8 & Shock & 46 & 0.23 & 0.02 & 0.31 & 0.31 \\
\hline
\end{tabular}
\label{t:normfits}
\end{table}

\section{Discussion}

\subsection{Spot size}

The apparent spot size at infinity $R_\mathrm{spot}$  can be related to the actual (circular) spot radius
by taking into account relativistic light bending, as described in \citet{2003MNRAS.343.1301P} and \citet{2009MNRAS.400..492I}.\footnote{Note that the relation is derived for a slowly rotating pulsar and blackbody emission.}
We can use the emitting areas shown in Table \ref{t:fits} for the group T to obtain estimations of the spot size.
We take a star with  \mstar=1.4 \msun\ and \rstar= 10.3 km (2.5  Schwarzschild radii).
The actual spot radius depends on the system inclination $i$ and the spot colatitude $\theta$ (see sect. 5.1 of \citealt{2009MNRAS.400..492I}).
The smallest and the largest possible radii correspond to  $i=\theta=0\degr$ and $i=\theta=90\degr$, respectively.
For best-fitting normalizations of the \xte-only fits, the interval of radii is 4.5--7.1 km, and for the joint \xte\ and \swift\ spectrum it is 4.0--6.4 km.

Besides the uncertainty in the areas because of a weakly constrained distance to the object, the numbers quoted are subjected to uncertainty that resides in the unknown relation between the emitting area of the blackbody and Comptonization components (which we assumed to be equal).
For SAX J1808.4--3658, the time-average spectrum (for the slow decay outburst stage) suggested that the blackbody area is twice as large as the Comptonized one, and the likely error in the area is about 50 per cent; additional  uncertainty due to a colour-correction appears if the emission is different from a blackbody, however, for atmospheres heated from above this correction should not play a significant role  \citep{2003MNRAS.343.1301P,2009MNRAS.400..492I}.  Finally, the estimation is valid for the filled circular spot, while in reality the spot shape can be rather different (see  Section \ref{s:ampl}).

\subsection{Oscillation amplitude and geometry}
\label{s:ampl}

\begin{figure}
\centerline{\epsfig{file=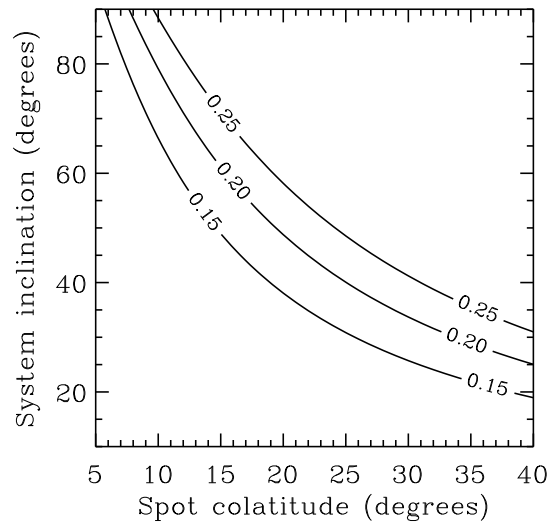,width=7cm}}
\caption{Contour plots of the constant oscillation amplitude at the plane inclination -- spot colatitude.
The curves are computed using expression (22) in \citet{2003MNRAS.343.1301P} (see also Sect 5.1 in \citealt{2009MNRAS.400..492I})
assuming a 1.4\msun\ neutron star with radius of 10.3 km and the spot size of $R_{\rm \infty}=5$ km.
}
\label{f:ampl}
\end{figure}

\citet{2003MNRAS.343.1301P} have derived an expression that relates  oscillation amplitude of the pulsar to the system inclination and spot colatitude for a case of large, blackbody-emitting, always visible spot on the surface of a slowly rotating star (see equation (10) there and sect. 5.2 in \citealt{2009MNRAS.400..492I}). While our hotspot is non-blackbody and the star rotates rapidly (and there are arguments against filled circular spot shape, see below), we can compare the observed relative amplitude of the fundamental  with this analytical relation to obtain a zero-order estimate of system's geometrical parameters.
Taking the apparent spot radius $R_{\rm \infty}=5$ km, we can obtain a dependence of the amplitude on the inclination and spot colatitude
which  (and typical neutron star parameters) are shown on Fig. \ref{f:ampl}. We remind that in our case  the observed amplitude of the fundamental is about 22 per cent at the beginning of the outburst, gradually decreasing to 15 per cent (Fig. \ref{f:sinefits}). Assuming 60\degr\ inclination
we get  the spot colatitude of about 15\degr. A decrease in the spot colatitude can cause a corresponding decrease of the amplitude.

What could be a reason for decreasing   spot colatitude during the outburst?
Present understanding of the AMP geometry is that of the inclined dipole with the accretion disc disrupted at some truncation radius
(likely proportional to the Alfv{\'e}n radius, see e.g. \citealt{2005ApJ...634.1214L}) by the magnetic field.
Matter flows along the magnetic field lines and falls on the neutron star surface forming a hotspot.
As the accretion rate drops, the Alfv{\'e}n radius increases and the disc recedes from the neutron star.
The matter then accretes along the magnetic field lines that touch the star closer to the magnetic pole
and thus hotspot outer radius decreases.
In case of a hypothetical circular spot, decreasing of the spot size $\rho$ would result in \textit{increasing} of the amplitude
(expression 22 in \citealt{2003MNRAS.343.1301P}), which contradicts the observations.
A filled circular spot, however, is not what is seen in MHD simulations, instead a ring- or a crescent-shaped spot shapes are observed \citep{2004ApJ...610..920R}.
In such a case, the emission is generated in a preferred sector situated closer to the disc.
When the outer spot radius decreases, the preferred sector shifts closer to the magnetic pole of the star
reducing the effective spot colatitude and leading to a decrease of the pulse amplitude as observed.

\subsection{Origin of the lag evolution}
\label{s:lags}

The phase lags can be explained by a difference in emissivity patterns at different energies
\citep{2002MNRAS.331..141G,2003MNRAS.343.1301P}.
At relatively high energies (above 10 keV) Comptonization is dominating the emission. But as we proceed downwards from 10 to few keV, we observe gradually increasing contribution from the blackbody component which alters the emissivity pattern (towards more isotropic emission). Therefore maximum of flux is observed  for different energies at different pulse phases and this effect creates energy-dependent time lags.
In agreement with this explanation, the observed lag saturates at the energy where blackbody component becomes insignificant.
\igr\ demonstrates a smooth increase in the time lags of the fundamental during the outburst  (Fig. \ref{f:lags})
and a nearly constant time lags of the overtone (rather weakly constrained by observations).
Several sources show that the lag  does not remain constant during the outburst: in SAX J1808.4--3658
the lag value  increases  during the ``slow decay", while in the ``flaring tail" stage it starts to decrease \citep{2009ApJ...697.2102H}.
In XTE J1814--338 there is also an observational evidence of the lag increasing in the end of the outburst \citep{2006MNRAS.373..769W}.

While the detailed modelling of actual pulse profiles is beyond the scope of this work, below we discuss the likely cause of the observed lag evolution.
The pulse shape depends on many factors, such as the spot shape (which can be different for different emission components) and size,
colatitude of the magnetic pole and the inner radius of accretion disc  \citep[see][]{2008AIPC.1068...77P}.
Changes in these parameters will affect the pulse shape, but not every of them can generate the observed increase of lags.

To study the lag evolution, we employed the following toy model (following the framework developed by \citealt{2003MNRAS.343.1301P,2004A&A...426..985V,  2006MNRAS.373..836P,2009MNRAS.400..492I}).
We have assumed a system inclination of 60\degr\ and a 1.4 \msun\ star with radius of 10.3 km  rotating at the pulsar frequency of 245 Hz.
To model the emission anisotropy, we have chosen the angular dependence of emitting radiation in the form $I(\alpha)=I_0(1-h \cos \alpha)$, where
$h$ is the anisotropy parameter and $\alpha$ is the angle between the spot normal and photon direction in the spot frame.
For the blackbody radiation $h=0$, while for Comptonization in a slab of Thomson optical depth about unity $h\gtrsim0.5$
\citep*{2004A&A...426..985V,2007MNRAS.381..723I}, and for the observed AMP pulse profiles $h\sim0.7$--$0.8$
is required \citep{2003MNRAS.343.1301P,2009ApJ...706L.129P}.
We generate two pulse profiles: one  in the hard energy band, where only the Comptonization component contributes, and
another at soft energies is a mixture of  the blackbody and  Comptonization components in equal proportions
(we neglect changes of the Comptonization emissivity pattern with energy, which is a good approximation, see \citealt{2007MNRAS.381..723I}).
We consider two geometries: a circular spot and a crescent spot, mimicking the shape obtained in  MHD modelling \citep{2004ApJ...610..920R}.

\begin{figure}
\centerline{\epsfig{file=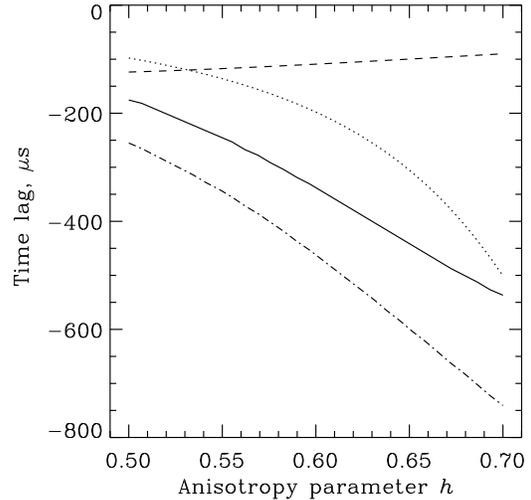,width=7cm}}
\caption{Time lag dependence on the anisotropy parameter $h$ from 0.5 to 0.7. The pulse maximum lags and
the lags of the fundamental and the overtone are shown as the solid, dotted and dashed curves, respectively.
The dash-dotted line represents the pulse maximum lag between the blackbody and the Comptonization component lightcurves.
The contribution of two components to the soft X-ray band are assumed to be equal, the spot colatitude $\theta$ was set to 15\degr,
the powerlaw photon index of  the continuum spectrum was assumed to be 1.9, and the angular radii of blackbody and Comptonization emitting spots were assumed  30\degr and 15\degr, respectively.
}
\label{f:modellag}
\end{figure}

For both assumed spot geometries, we have arrived to a similar conclusion.
A rather small change of anisotropy parameter (from 0.5 to 0.6--0.7) turns out to be the best candidate for the lag evolution, as it can reproduce the observed lag increase of about $\sim200\mu$s  (see Fig. \ref{f:modellag}).
On the other hand, a physically realistic change of the ratio between Comptonized and blackbody emission in the ``soft" light curve, a change of the relative size of Comptonization and blackbody spot, a change in the spot size or
a change of the effective spot colatitude were not able to produce lag evolution large enough to match the observed value.


Additionally, we can compare the aforementioned time lags with the lags between \textit{physical} components (blackbody and Comptonization light curves). Very similarly to what is observed with the phase-resolved spectroscopy,
we obtain  the ``component lags" that are nearly two times bigger than the lags computed as a mixture of two components. For the case shown on Fig. \ref{f:modellag}, the ``component lag'' changes from $-$250 to $-$740 $\mu$s.

Variation of $h$ in principle could be related to a decreasing optical depth of the Comptonizing slab (see Table \ref{t:fits}),
which would allow for more blackbody emission to contribute to the soft energy band. A decrease of $\tau_{\rm T}$ from 1.85 to 1.74 would
result in a $\sim$10  per cent larger blackbody flux, making  it unlikely to account for the observed increase of the phase lag.
To achieve a lag increase of 200 $\mu$s for a typical case shown in Fig. \ref{f:modellag}, the ratio  of the blackbody flux to the Comptonized one in the soft band should change by a factor of several, which contradicts the observations of a remarkably similar spectra seen during the outburst.
This would mean that the intrinsic angular emissivity of emitting plasma should not change much.
However, the vertical structure of the accretion shock might change with variations of the accretion rate causing
changes in the emissivity pattern.
Thus, if the anisotropy parameter $h$ is varying during the outburst, it might indicate variations in
the accretion shock structure.
Alternatively,
a physical displacement of the centroid of the Comptonization component (accretion shock) relative  to the blackbody emitting spot
can cause the lag increase.


\section{Conclusions}

In this paper, we have analysed the data from \igr\ obtained by \xte\ and \swift\ during its  September--October 2009 outburst.
For the largest part of the outburst the source was in the ``slow decay" outburst stage.
However, after 24 days from the beginning of outburst, right before the source faded beyond \xte\ detection limit (MJD 55112),
a very sharp decrease of the source flux was detected accompanied by a considerable phase shift of the pulse profile.
This clearly indicates the end of the ``slow decay" stage at that date.

The energy spectra of \igr\ can be well described with Comptonization in the accretion shock ($\kte \sim 30$ keV, $\tau_{\rm T} \sim 2$) and soft $\sim$1 keV blackbody emission originating from the NS surface.
The \swift/XRT data allowed us to estimate the accretion disc parameters.
We obtained the inner disc temperature of about 0.24  keV
and reasonable inner disc radius estimates located within the corotation radius.
We also detected weak reflection and an iron line ($EW\sim60$--$100$ eV). Spectral evolution was rather weak during the outburst. Only in the very end of the outburst (after MJD 55105), we detected a slight change in the optical depth $\tau_{\rm T}$ of the Comptonized emission; such a spectral stability is known to be common for AMPs.

The pulse profiles of the source were smooth and do not demonstrate any prominent narrow secondary features, which indicates that observed emission comes mainly from one emitting spot. The pulse harmonics demonstrate drop of their amplitudes during the course of outburst.
The study of phase lag behaviour revealed a considerable increase of the lag from 150 to 400 $\mu$s, an effect seen also  in few other AMPs.
A change in the anisotropy pattern  of the Comptonized radiation could explain the observed time lag evolution,
implying  changes in the accretion column geometry or a physical displacement of the centroid of the accretion shock relative  to the blackbody spot.

\section*{Acknowledgements}
We thank Alessandro Patruno for helpful discussions. AI was supported by EU FP6 Transfer of Knowledge Project ``Astrophysics of Neutron Stars" (MTKD-CT-2006-042722).
JJEK acknowledges the Finnish Graduate School in Astronomy and Space Physics and JP the Academy of Finland grant 127512.


\begin{thebibliography}{53}
\expandafter\ifx\csname natexlab\endcsname\relax\def\natexlab#1{#1}\fi

\bibitem[{{Altamirano} {et~al}\mbox{.}(2010){Altamirano}, {Watts}, {Linares},
  {Markwardt}, {Strohmayer}, \& {Patruno}}]{2010MNRAS.tmp.1363A}
{Altamirano} D., {Watts} A., {Linares} M., {Markwardt} C.~B., {Strohmayer} T.,
  {Patruno} A., 2010, \mnras, 1363

\bibitem[{{Arnaud}(1996)}]{1996ASPC..101...17A}
{Arnaud} K.~A.,  1996, in {Jacoby} G.~H.,  {Barnes} J.,  eds, ASP Conf. Ser. Vol. 101,
Astronomical Data  Analysis Software and Systems V.
Astron. Soc. Pac., San Francisco, p. 17

\bibitem[{{Baldovin} {et~al}\mbox{.}(2009){Baldovin}, {Kuulkers}, {Ferrigno},
  {Bozzo}, {Chenevez}, {Brandt}, {Beckmann}, {Bird}, {Domingo}, {Ebisawa},
  {Jonker}, {Kretschmar}, {Markwardt}, {Oosterbroek}, {Paizis}, {Risquez},
  {Sanchez-Fernandez}, {Shaw}, \& {Wijnands}}]{2009ATel.2196....1B}
{Baldovin} C. {et~al.}, 2009, The Astronomer's Telegram, 2196, 1

\bibitem[{{Bozzo} {et~al}\mbox{.}(2010){Bozzo}, {Ferrigno}, {Falanga},
  {Campana}, {Kennea}, \& {Papitto}}]{2010A&A...509L...3B}
{Bozzo} E., {Ferrigno} C., {Falanga} M., {Campana} S., {Kennea} J.~A.,
  {Papitto} A., 2010, \aap, 509, L3

\bibitem[{{Bozzo} {et~al}\mbox{.}(2009){Bozzo}, {Ferrigno}, {Kuulkers},
  {Falanga}, {Chenevez}, {Brandt}, {Beckmann}, {Bird}, {Domingo}, {Ebisawa},
  {Jonker}, {Kretschmar}, {Markwardt}, {Oosterbroek}, {Paizis}, {Risquez},
  {Sanchez-Fernandez}, {Shaw}, \& {Wijnands}}]{2009ATel.2198....1B}
{Bozzo} E. {et~al.}, 2009, The Astronomer's Telegram, 2198, 1

\bibitem[{{Cui} {et~al}\mbox{.}(1998){Cui}, {Morgan}, \&
  {Titarchuk}}]{1998ApJ...504L..27C}
{Cui} W., {Morgan} E.~H., {Titarchuk} L.~G., 1998, \apjl, 504, L27

\bibitem[{{Done} {et~al}\mbox{.}(2007){Done}, {Gierli{\'n}ski},
  \& {Kubota}}]{2007A&ARv..15....1D}
{Done} C., {Gierli{\'n}ski} M., {Kubota} A., 2007, \aapr, 15, 1

\bibitem[{{Evans} {et~al}\mbox{.}(2009){Evans} {et al.}}]{2009MNRAS.397.1177E}
{Evans} P. A. {et~al.}, 2009,  MNRAS, 397, 1177

\bibitem[{{Falanga} {et~al}\mbox{.}(2005{\natexlab{a}}){Falanga},
  {Bonnet-Bidaud}, {Poutanen}, {Farinelli}, {Martocchia}, {Goldoni}, {Qu},
  {Kuiper}, \& {Goldwurm}}]{2005A&A...436..647F}
{Falanga} M. {et~al.}, 2005{\natexlab{a}}, \aap, 436, 647

\bibitem[{{Falanga} {et~al}\mbox{.}(2005{\natexlab{b}}){Falanga}, {Kuiper},
  {Poutanen}, {Bonning}, {Hermsen}, {di Salvo}, {Goldoni}, {Goldwurm}, {Shaw},
  \& {Stella}}]{2005A&A...444...15F}
---, 2005{\natexlab{b}}, \aap, 444, 15

\bibitem[{{Falanga} {et~al}\mbox{.}(2011){Falanga}, {Kuiper}, {Poutanen},
  {Galloway}, {Bonning}, {Bozzo}, {Goldwurm}, {Hermsen}, \&
  {Stella}}]{2010arXiv1012.0229F}
---, 2011, \aap, in press [arXiv:astro-ph/1012.0229]

\bibitem[{{Falanga} {et~al}\mbox{.}(2007){Falanga}, {Poutanen}, {Bonning},
  {Kuiper}, {Bonnet-Bidaud}, {Goldwurm}, {Hermsen}, \&
  {Stella}}]{2007A&A...464.1069F}
{Falanga} M., {Poutanen} J., {Bonning} E.~W., {Kuiper} L., {Bonnet-Bidaud}
  J.~M., {Goldwurm} A., {Hermsen} W., {Stella} L., 2007, \aap, 464, 1069

\bibitem[{{Frank}, {King} \& {Raine}(2002){Frank}, {King}, \& {Raine}}]{FKR02}
{Frank} J., {King} A., {Raine} D.~J., 2002, {Accretion Power in Astrophysics}.
  Cambridge University Press, Cambridge

\bibitem[{{Galloway} {et~al}\mbox{.}(2005){Galloway}, {Markwardt}, {Morgan},
  {Chakrabarty}, \& {Strohmayer}}]{2005ApJ...622L..45G}
{Galloway} D.~K., {Markwardt} C.~B., {Morgan} E.~H., {Chakrabarty} D.,
  {Strohmayer} T.~E., 2005, \apjl, 622, L45

\bibitem[{{Galloway} {et~al}\mbox{.}(2008){Galloway}, {Muno}, {Hartman},
  {Psaltis}, \& {Chakrabarty}}]{2008ApJS..179..360G}
{Galloway} D.~K., {Muno} M.~P., {Hartman} J.~M., {Psaltis} D., {Chakrabarty}
  D., 2008, \apjs, 179, 360

\bibitem[{{Gierli{\'n}ski} {et~al}\mbox{.}(2002){Gierli{\'n}ski}, {Done},
  \& {Barret}}]{2002MNRAS.331..141G}
{Gierli{\'n}ski} M., {Done} C., {Barret} D., 2002, \mnras, 331, 141

\bibitem[{{Gierli{\'n}ski} \& {Poutanen}(2005)}]{2005MNRAS.359.1261G}
{Gierli{\'n}ski} M., {Poutanen} J., 2005, \mnras, 359, 1261

\bibitem[{{Gierli{\'n}ski} {et~al}\mbox{.}(1999){Gierli{\'n}ski}, {Zdziarski},
  {Poutanen}, {Coppi}, {Ebisawa}, \& {Johnson}}]{1999MNRAS.309..496G}
{Gierli{\'n}ski} M., {Zdziarski} A.~A., {Poutanen} J., {Coppi} P.~S., {Ebisawa}
  K., {Johnson} W.~N., 1999, \mnras, 309, 496

\bibitem[{{Hartman} {et~al}\mbox{.}(2008){Hartman}, {Patruno}, {Chakrabarty},
  {Kaplan}, {Markwardt}, {Morgan}, {Ray}, {van der Klis}, \&
  {Wijnands}}]{2008ApJ...675.1468H}
{Hartman} J.~M. {et~al.}, 2008, \apj, 675, 1468

\bibitem[{{Hartman} {et~al}\mbox{.}(2009){Hartman}, {Watts}, \&
  {Chakrabarty}}]{2009ApJ...697.2102H}
{Hartman} J.~M., {Watts} A.~L., {Chakrabarty} D., 2009, \apj, 697, 2102

\bibitem[{{Ibragimov} \& {Poutanen}(2009)}]{2009MNRAS.400..492I}
{Ibragimov} A., {Poutanen} J., 2009, \mnras, 400, 492

\bibitem[{{Ibragimov} {et~al}\mbox{.}(2007){Ibragimov}, {Zdziarski},
  \& {Poutanen}}]{2007MNRAS.381..723I}
{Ibragimov} A., {Zdziarski} A.~A., {Poutanen} J., 2007, \mnras, 381, 723

\bibitem[{{Jahoda} {et~al}\mbox{.}(2006){Jahoda}, {Markwardt}, {Radeva},
  {Rots}, {Stark}, {Swank}, {Strohmayer}, \& {Zhang}}]{2006ApJS..163..401J}
{Jahoda} K., {Markwardt} C.~B., {Radeva} Y., {Rots} A.~H., {Stark} M.~J.,
  {Swank} J.~H., {Strohmayer} T.~E., {Zhang} W., 2006, \apjs, 163, 401

\bibitem[{{Kubota} {et~al}\mbox{.}(1998){Kubota}, {Tanaka}, {Makishima},
  {Ueda}, {Dotani}, {Inoue}, \& {Yamaoka}}]{1998PASJ...50..667K}
{Kubota} A., {Tanaka} Y., {Makishima} K., {Ueda} Y., {Dotani} T., {Inoue} H.,
  {Yamaoka} K., 1998, \pasj, 50, 667

\bibitem[{{Kuulkers} {et~al}\mbox{.}(2003){Kuulkers}, {den Hartog}, {in't
  Zand}, {Verbunt}, {Harris}, \& {Cocchi}}]{2003A&A...399..663K}
{Kuulkers} E., {den Hartog} P.~R., {in't Zand} J.~J.~M., {Verbunt} F.~W.~M.,
  {Harris} W.~E., {Cocchi} M., 2003, \aap, 399, 663

\bibitem[{{Kuulkers} {et~al}\mbox{.}(2007){Kuulkers}, {Shaw}, {Paizis},
  {Chenevez}, {Brandt}, {Courvoisier}, {Domingo}, {Ebisawa}, {Kretschmar},
  {Markwardt}, {Mowlavi}, {Oosterbroek}, {Orr}, {R{\'{\i}}squez},
  {Sanchez-Fernandez}, \& {Wijnands}}]{2007A&A...466..595K}
{Kuulkers} E. {et~al.}, 2007, \aap, 466, 595


\bibitem[{{Leahy} {et~al}\mbox{.}(2008){Leahy}, {Morsink}, \&
  {Cadeau}}]{2008ApJ...672.1119L}
{Leahy} D.~A., {Morsink} S.~M., {Cadeau} C., 2008, \apj, 672, 1119

\bibitem[{{Long} {et~al}\mbox{.}(2005){Long}, {Romanova}, \&
  {Lovelace}}]{2005ApJ...634.1214L}
{Long} M., {Romanova} M.~M., {Lovelace} R.~V.~E., 2005, \apj, 634, 1214

\bibitem[{{Magdziarz} \& {Zdziarski}(1995)}]{1995MNRAS.273..837M}
{Magdziarz} P., {Zdziarski} A.~A., 1995, \mnras, 273, 837

\bibitem[{{Markwardt} {et~al}\mbox{.}(2009{\natexlab{a}}){Markwardt},
  {Altamirano}, {Strohmayer}, \& {Swank}}]{2009ATel.2237....1M}
{Markwardt} C.~B., {Altamirano} D., {Strohmayer} T.~E., {Swank} J.~H.,
  2009{\natexlab{a}}, The Astronomer's Telegram, 2237, 1

\bibitem[{{Markwardt} {et~al}\mbox{.}(2009{\natexlab{b}}){Markwardt},
  {Altamirano}, {Swank}, {Strohmayer}, {Linares}, \&
  {Pereira}}]{2009ATel.2197....1M}
{Markwardt} C.~B., {Altamirano} D., {Swank} J.~H., {Strohmayer} T.~E.,
  {Linares} M., {Pereira} D., 2009{\natexlab{b}}, The Astronomer's Telegram,
  2197, 1

\bibitem[{{Miller-Jones} {et~al}\mbox{.}(2009){Miller-Jones},
  {Russell}, \& {Migliari}}]{2009ATel.2232....1M}
{Miller-Jones} J.~C.~A., {Russell} D.~M., {Migliari} S., 2009, The Astronomer's
  Telegram, 2232, 1

\bibitem[{{Nowak} {et~al}\mbox{.}(2009){Nowak}, {Paizis}, {Wilms}, {Rodriguez},
  {Chaty}, {Ebisawa}, {Del Santo}, {Farinelli}, {Ubertini}, \&
  {Courvoisier}}]{2009ATel.2215....1N}
{Nowak} M.~A. {et~al.}, 2009, The Astronomer's Telegram, 2215, 1

\bibitem[{{Papitto} {et~al}\mbox{.}(2010){Papitto}, {Riggio}, {di Salvo},
  {Burderi}, {D'A{\`i}}, {Iaria}, {Bozzo}, \& {Menna}}]{2010MNRAS.407.2575P}
{Papitto} A., {Riggio} A., {di Salvo} T., {Burderi} L., {D'A{\`i}} A., {Iaria}
  R., {Bozzo} E., {Menna} M.~T., 2010, \mnras, 407, 2575

\bibitem[{{Patruno}(2010)}]{2010arXiv1007.1108P}
{Patruno} A., 2010, PoS(HTRA-IV)028 [arXiv:astro-ph/1007.1108]

\bibitem[{{Patruno} {et~al}\mbox{.}(2009){Patruno}, {Rea}, {Altamirano},
  {Linares}, {Wijnands}, \& {van der Klis}}]{2009MNRAS.396L..51P}
{Patruno} A., {Rea} N., {Altamirano} D., {Linares} M., {Wijnands} R., {van der
  Klis} M., 2009, \mnras, 396, L51


\bibitem[{{Poutanen}(2006)}]{P06}
{Poutanen} J.,  2006, Adv. Space Res., 38, 2697

\bibitem[{{Poutanen}(2008)}]{2008AIPC.1068...77P}
{Poutanen} J.,  2008, in {Wijnands} R.,  {Altamirano} D.,  {Soleri} P.,
  {Degenaar} N.,  {Rea} N.,  {Casella} P.,  {Patruno} A.,   {Linares} M.,  eds,
AIP Conf. Proc. Vol. 1068,
A Decade of Accreting X-ray Millisecond Pulsars, Am. Inst. Phys., New York, p. 77

\bibitem[{{Poutanen} \& {Beloborodov}(2006)}]{2006MNRAS.373..836P}
{Poutanen} J., {Beloborodov} A.~M., 2006, \mnras, 373, 836

\bibitem[{{Poutanen} \& {Gierli{\'n}ski}(2003)}]{2003MNRAS.343.1301P}
{Poutanen} J., {Gierli{\'n}ski} M., 2003, \mnras, 343, 1301

\bibitem[{{Poutanen} {et~al}\mbox{.}(2009){Poutanen}, {Ibragimov}, \&
  {Annala}}]{2009ApJ...706L.129P}
{Poutanen} J., {Ibragimov} A., {Annala} M., 2009, \apjl, 706, L129

\bibitem[{{Poutanen} \& {Svensson}(1996)}]{1996ApJ...470..249P}
{Poutanen} J., {Svensson} R., 1996, \apj, 470, 249

\bibitem[{{Revnivtsev} {et~al}\mbox{.}(2009){Revnivtsev}, {Sazonov},
  {Churazov}, {Forman}, {Vikhlinin}, \& {Sunyaev}}]{2009Natur.458.1142R}
{Revnivtsev} M., {Sazonov} S., {Churazov} E., {Forman} W., {Vikhlinin} A.,
  {Sunyaev} R., 2009, \nat, 458, 1142

\bibitem[{{Riggio} {et~al}\mbox{.}(2011){Riggio}, {Papitto}, {Burderi}, {Di
  Salvo}, {Bachetti}, {Iaria}, {D'A{\`i}}, \& {Menna}}]{2011A&A...526A..95R}
{Riggio} A., {Papitto} A., {Burderi} L., {Di Salvo} T., {Bachetti} M., {Iaria}
  R., {D'A{\`i}} A., {Menna} M.~T., 2011, A\&A,  526, A9

\bibitem[{{Riggio} {et~al}\mbox{.}(2009){Riggio}, {Papitto}, {Burderi}, {di
  Salvo}, {D'Ai}, {Iaria}, \& {Menna}}]{2009ATel.2221....1R}
{Riggio} A., {Papitto} A., {Burderi} L., {di Salvo} T., {D'Ai} A., {Iaria} R.,
  {Menna} M.~T., 2009, The Astronomer's Telegram, 2221, 1

\bibitem[{{Romanova} {et~al}\mbox{.}(2004){Romanova}, {Ustyugova}, {Koldoba},
  \& {Lovelace}}]{2004ApJ...610..920R}
{Romanova} M.~M., {Ustyugova} G.~V., {Koldoba} A.~V., {Lovelace} R.~V.~E.,
  2004, \apj, 610, 920

\bibitem[{{Shimura} \& {Takahara}(1995)}]{1995ApJ...445..780S}
{Shimura} T., {Takahara} F., 1995, \apj, 445, 780

\bibitem[{{Torres} {et~al}\mbox{.}(2009){Torres}, {Jonker}, {Steeghs},
  {Damjanov}, {Caris}, \& {Glazebrook}}]{2009ATel.2233....1T}
{Torres} M.~A.~P., {Jonker} P.~G., {Steeghs} D., {Damjanov} I., {Caris} E.,
  {Glazebrook} K., 2009, The Astronomer's Telegram, 2233, 1

\bibitem[{{Tsujimoto} {et~al}\mbox{.}(2011){Tsujimoto}, {Guainazzi},
  {Plucinsky}, {Beardmore}, {Ishida}, {Natalucci}, {Posson-Brown}, {Read},
  {Saxton}, \& {Shaposhnikov}}]{2011A&A...525A..25T}
{Tsujimoto} M. {et~al.}, 2011, A\&A, 525, A25

\bibitem[{{Viironen} \& {Poutanen}(2004)}]{2004A&A...426..985V}
{Viironen} K., {Poutanen} J., 2004, \aap, 426, 985

\bibitem[{{Watts} {et~al}\mbox{.}(2009){Watts}, {Altamirano}, {Markwardt}, \&
  {Strohmayer}}]{2009ATel.2199....1W}
{Watts} A.~L., {Altamirano} D., {Markwardt} C.~B., {Strohmayer} T.~E., 2009,
  The Astronomer's Telegram, 2199, 1

\bibitem[{{Watts} \& {Strohmayer}(2006)}]{2006MNRAS.373..769W}
{Watts} A.~L., {Strohmayer} T.~E., 2006, \mnras, 373, 769

\bibitem[{{Wijnands}(2006)}]{W06}
Wijnands R., 2006, in Lowry  J. A., ed., Trends in Pulsar Research.
Nova Science Publishers, New York, p. 53

\bibitem[{{Wilkinson} {et~al}\mbox{.}(2010){Wilkinson}, {Patruno}, {Watts}, \&
  {Uttley}}]{2010MNRAS.tmp.1560W}
{Wilkinson} T., {Patruno} A., {Watts} A., {Uttley} P., 2010, \mnras, 1560



\end{thebibliography}

\label{lastpage}
\end{document}